\documentclass[prc,twocolumn,superscriptaddress,showpacs,psfig,showkeys,nofootinbib]{revtex4}
\usepackage{amsmath}
\usepackage{amssymb}
\usepackage{amsfonts}
\usepackage{graphicx}
\usepackage{epsfig}
\usepackage{dcolumn}
\usepackage{hyperref}
\usepackage{epstopdf}
\usepackage{color}
\usepackage[utf8]{inputenc}
\usepackage{amsthm}
\usepackage{fontenc}
\usepackage{bm}
\RequirePackage{slashed}\usepackage{cancel}
\usepackage[normalem]{ulem}

\usepackage{array,booktabs}
\newcolumntype{C}{>{$\displaystyle}c<{$}}

\usepackage{soul}

\begin{document}


\title{Photoproduction of $J/\psi$ in d+Au ultraperipheral collisions at the BNL Relativistic Heavy Ion Collider}

\author{V. Guzey}
\affiliation{National Research Center ``Kurchatov Institute'', Petersburg Nuclear Physics Institute (PNPI), Gatchina, 188300, Russia}
\author{E. Kryshen}
\affiliation{National Research Center ``Kurchatov Institute'', Petersburg Nuclear Physics Institute (PNPI), Gatchina, 188300, Russia}
\author{M. Strikman}
\affiliation{Pennsylvania  State  University,  University  Park,  PA,  16802,  USA} 
\author{M. Zhalov}
\affiliation{National Research Center ``Kurchatov Institute'', Petersburg Nuclear Physics Institute (PNPI), Gatchina, 188300, Russia}

\date{\today}

\begin{abstract}

We argue that the STAR data on $J/\psi$ photoproduction in $d+Au$ ultraperipheral collisions (UPCs) 
at the momentum transfer $t \neq 0$ primarily probes elastic and possibly nucleon-dissociative $J/\psi$ photoproduction on quasi-free nucleons, 
while being essentially insensitive to expectedly weak nuclear modifications of the gluon distribution in the deuteron. 
Analyzing the STAR detector capabilities for selection of UPC events, 
we explore the possibility that the contribution of
nucleon-dissociative $J/\psi$ photoproduction is negligibly small 
and obtain a very good description of the $t$ dependence of the STAR total cross section.

\end{abstract}

\maketitle

\section{Introduction}
\label{sec:intro}

Ultraperipheral collisions (UPCs) of relativistic ions have become a standard tool for studies of quantum chromodynamics 
(QCD)~\cite{Baltz:2007kq,Klein:2019qfb}. In particular, photoproduction of $J/\psi$ vector mesons off heavy nuclear targets in 
Pb-Pb UPCs at the Large Hadron Collider (LHC) provided new constraints on the nuclear gluon distribution at small $x$, 
see Ref.~\cite{Guzey:2020ntc} for references to the experimental data. These results have been interpreted within
 the leading twist approximation (LTA)~\cite{Guzey:2013xba,Guzey:2013qza,Guzey:2020ntc}
and also in terms of proton shape fluctuations in the context of the color glass condensate (CGC) framework and the color dipole model~\cite{Mantysaari:2017dwh,Mantysaari:2020axf}.

Recently the STAR collaboration at Brookhaven National Laboratory (BNL) for the first time measured $J/\psi$ photoproduction in $d+Au$ UPCs at $\sqrt{s_{NN}}=200$ GeV~\cite{STAR:2021wwq}. Since the flux of equivalent photons from ions of Au is much larger 
than that from the deuteron (the photon flux scales as $Z^2$ with $Z$ being the ion electric charge),
these data are treated as photoproduction off the deuteron target. The results for the differential cross section as a function of the momentum transfer squared $t$ were compared in~\cite{STAR:2021wwq} to theoretical predictions of the two approaches mentioned above and it was found that the CGC saturation model
provides a better agreement, primarily at large $|t|$. Note that the theoretical predictions were made for generic RHIC kinematics
and without taking into account the specifics of STAR detector capabilities for selection of UPC events.

The purpose of this 
paper
 is to demonstrate that since the STAR $d+Au$ UPC data corresponds to the photon-nucleon invariant energy $W \approx 25$ GeV and, hence, to the parton momentum fraction $x \approx 0.015$,  
nuclear modifications of the gluon distribution in the deuteron expectedly play an insignificant 
role in interpretation of the data, see also the discussion in~\cite{STAR:2021wwq}; the enhancement in the $t\to 0$ limit
is well-understood and explained by nuclear coherence of the gluon density of the deuteron.
Instead, the data at $t \neq 0$ is sensitive primarily to elastic and 
possibly nucleon-dissociative $J/\psi$ photoproduction on quasi-free nucleons, which were previously studied at HERA at high $W$.
In particular, we explore the scenario that the dominant range of the produced nucleon-dissociative masses and the corresponding 
rapidity interval were vetoed by the STAR detector system selecting the UPC events, 
set the nucleon-dissociative contribution to zero, and obtain a very good agreement with the STAR data on the total cross section
over the entire range of $t$.
We also suggest 
that the difference between the LTA and CGC predictions 
could come from different extrapolations of the  
 elastic and nucleon-dissociative $J/\psi$ photoproduction cross sections on the nucleon target from HERA to RHIC energies.

The remainder of the paper is organized as follows. In Sec.~\ref{sec:summed}, within the LTA, we present the derivation of the cross section of $J/\psi$ photoproduction
off the deuteron, when the nuclear target can stay intact or dissociate (so-called summed cross section).
Our predictions for the $t$ dependence of the $d+Au$ UPC cross section and their comparison to the STAR data are discussed in Sec.~\ref{sec:lta}.
Finally, we summarize our findings in Sec.~\ref{sec:summary}.

%
\section{Summed cross section off the deuteron}
\label{sec:summed}

Applying the formalism of Ref.~\cite{Guzey:2018tlk} to $J/\psi$ photoproduction on the deuteron
and using completeness of final nuclear states,
one obtains the following expression for the summed (coherent+incoherent) $\gamma +d \to J/\psi+X$ differential cross section
\begin{eqnarray}
&&\frac{d\sigma_d^{\rm sum}(t)}{dt}= \nonumber\\
&&\frac{\chi^2}{4\pi} \int d^2 {\vec b} \int d^2 {\vec b^{\prime}} e^{i \vec{q} \cdot 
({\vec b}-\vec{b^{\prime}})} \langle 0|\Gamma_d^{\dagger}(\vec{b^{\prime}},\vec{r}) \Gamma_d(\vec{b},\vec{r}) |0 \rangle \,,
\label{eq:cs_sum}
\end{eqnarray}
where $\chi$ is proportional to the $\gamma - J/\psi$ transition amplitude.
In Eq.~(\ref{eq:cs_sum}),
$\Gamma_d({\vec b},\vec{r})$ is the scattering amplitude on the deuteron in impact parameter space; 
$\vec{b}$ and ${\vec q}$ are two-dimensional vectors of the impact parameter and the momentum transfer, respectively;
the latter is taken to be transverse, $t=-|{\vec q}|^2$.
The matrix element $\langle 0| \dots |0\rangle$ 
stands for the integration with the deuteron ground-state wave function $\Psi_d(\vec{r})$ squared
\begin{equation}
\langle 0| \Gamma_d^{\dagger}(\vec{b^{\prime}},\vec{r}) \Gamma_d(\vec{b},\vec{r}) |0\rangle=\int d^3 \vec{r}\, |\Psi_d(\vec{r})|^2 \Gamma_d^{\dagger}(\vec{b^{\prime}},\vec{r}) \Gamma_d(\vec{b},\vec{r})\,,
\label{eq:wf}
\end{equation}
where $\vec{r}$ is the relative distance between the proton and the neutron in the deuteron; averaging over 
the deuteron polarization is implied.
In our analysis, we used the Paris deuteron wave function~\cite{Lacombe:1981eg}.

In the Glauber multiple scattering approach~\cite{Bauer:1977iq},
the $\gamma +d \to J/\psi+X$ nuclear scattering amplitude $\Gamma_d(\vec{b},\vec{r})$
can be expressed in terms of 
the nucleon scattering amplitudes $\Gamma_N$ corresponding to the $\gamma +N \to J/\psi+N$ process (elastic production)
and $\Gamma_Y$ corresponding to the $\gamma +N \to J/\psi+Y$ process (nucleon dissociation), respectively,
as follows (we do not distinguish protons and neutrons)
\begin{eqnarray}
\Gamma_d(\vec{b},\vec{r})&=&\Gamma_N(\vec{b}-\frac{\vec r}{2})+\Gamma_N(\vec{b}+\frac{\vec r}{2})-\Gamma_N(\vec{b}-\frac{\vec r}{2})\Gamma_N(\vec{b}+\frac{\vec r}{2}) \nonumber\\
&+&\Gamma_Y(\vec{b}-\frac{\vec r}{2})+\Gamma_Y(\vec{b}+\frac{\vec r}{2})-\Gamma_N(\vec{b}-\frac{\vec r}{2})\Gamma_Y(\vec{b}+\frac{\vec r}{2}) \nonumber\\
&-& \Gamma_Y(\vec{b}-\frac{\vec r}{2})\Gamma_N(\vec{b}+\frac{\vec r}{2}) 
-\Gamma_Y(\vec{b}-\frac{\vec r}{2})\Gamma_Y(\vec{b}+\frac{\vec r}{2}) \,,
\label{eq:Gamma_d}
\end{eqnarray}
where $\vec{b}-\vec{r}/2$ and $\vec{b}+\vec{r}/2$ denote the coordinates of the two nucleons in the deuteron.
Equation~(\ref{eq:Gamma_d}) corresponds to the interaction with one and two nucleons of the nuclear target and
generalizes the classic result of the Gribov-Glauber theory of nuclear shadowing for hadron-deuteron scattering~\cite{Glauber:1955qq,Gribov:1968jf} by 
including the contribution of nucleon dissociation.

The nucleon scattering amplitudes are parametrized in the following form~\cite{Guzey:2018tlk}
\begin{eqnarray}
\Gamma_N(\vec{b}) &=& \frac{\sqrt{16 \pi d\sigma_{\rm el}(t=0)/dt}}{4 \pi B_{\rm el} \chi} e^{-|\vec{b}|^2/(2B_{\rm el})} \,, \nonumber\\
\Gamma_Y(\vec{b}) &=& \frac{\sqrt{16 \pi d\sigma_{\rm diss}(t=0)/dt}}{4 \pi \chi} \int \frac{d^2 \vec{q^{\prime}}}{2 \pi} e^{-i\vec{q^{\prime}} \cdot \vec{b}} f_{pd}(t^{\prime})\,,
\end{eqnarray}
where $d\sigma_{\rm el}(t)/dt$ is the cross section of the $\gamma + N \to J/\psi+N$ elastic production on the nucleon
and $B_{\rm el}=4.3 \pm 0.2$ GeV$^{-2}$~\cite{Alexa:2013xxa} is the slope of its $t$ dependence;
$d\sigma_{\rm diss}(t)/dt$ is the cross section of the nucleon-dissociative $J/\psi$ photoproduction $\gamma +N \to J/\psi+Y$ with
$f_{pd}(t)$ giving its $t$ dependence, see details below.

Substituting Eq.~(\ref{eq:Gamma_d}) in Eq.~(\ref{eq:cs_sum}), one notices that the resulting terms can be grouped together as follows
(we need to consider only the terms even in powers of $\Gamma_Y$ since the terms odd in $\Gamma_Y$ do not contribute to the matrix element
in Eq.~(\ref{eq:wf})).
The terms $\Gamma_N^{\dagger}(\vec{b^{\prime}} \pm \vec{r}/2)\Gamma_N(\vec{b} \pm \vec{r}/2)$ decouple from the integration with 
$|\Psi_d(\vec{r})|^2$ and
correspond to the $d\sigma_{\rm el}(t)/dt$ cross section,
while the terms 
$\Gamma_Y^{\dagger}(\vec{b^{\prime}} \pm \vec{r}/2)\Gamma_Y(\vec{b} \pm \vec{r}/2)$
 correspond to $d\sigma_{\rm diss}/dt$.
The interference terms $\Gamma_N^{\dagger}(\vec{b^{\prime}} \pm\vec{r}/2)\Gamma_N(\vec{b} \mp \vec{r}/2)$, when integrated with the 
deuteron wave function squared, result in the product of the deuteron charge form factor $F_d(4t)$ and 
the $d\sigma_{\rm el}(t)/dt$ cross section on the nucleon. Note that $4t$-argument of the deuteron form factor 
is a characteristic feature of the summed cross section
 and the nuclear shadowing correction, which originates from taking into account the effect of the deuteron center of mass.
A similar contribution results from the $\Gamma_Y^{\dagger}(\vec{b^{\prime}} \pm\vec{r}/2)\Gamma_Y(\vec{b} \mp \vec{r}/2)$ 
interference terms.

The higher terms ${\cal O}(\Gamma_N^3)$ give a small negative contribution to the summed cross section and, hence, describe 
the effect of nuclear shadowing. While their effect is negligibly small in the considered kinematics, it is still instructive to 
present the LTA shadowing correction to the summed nuclear cross section since it has not been discussed in the literature.
For a review and extensive references on early results on photoproduction of light and heavy vector mesons off light and heavy nuclei,
see Ref.~\cite{Bauer:1977iq}.
 To understand 
the structure of the answer, one notices that, e.g., in the $\Gamma_N^{\dagger}(\vec{b^{\prime}} - \vec{r}/2)\Gamma_N(\vec{b} - \vec{r}/2)\Gamma_N(\vec{b} + \vec{r}/2)$ term, the first factor is proportional to the elastic $\gamma+ N \to J/\psi+N$ amplitude, 
while the product of the last two factors is proportional to the $(d\sigma_{\rm diff}/dt)/\sigma_{\rm tot}$ ratio of 
the diffractive and total cross sections of inclusive deep inelastic scattering (DIS) on the nucleon $\gamma^{\ast} + N \to X +N$~\cite{Frankfurt:2011cs}.

Neglecting a vanishingly small contribution ${\cal O}(\Gamma_Y^2)$ to the shadowing term, one obtains the following expression for 
the summed cross section
\begin{eqnarray}
&&\frac{d\sigma_d^{\rm sum}(t)}{dt} =
2 \left(1+F_d(4 t)\right)\left(\frac{d\sigma_{\rm el}(t)}{dt}+\frac{d\sigma_{\rm diss}(t)}{dt}\right) \nonumber\\
&-&8\frac{d\sigma_{\rm el}(t=0)}{dt} e^{-B_{el}t/2} \nonumber\\
 & \times&  \frac{d\sigma_{\rm diff}/dt(t=0)}{\sigma_{\rm tot}}  \int \frac{d^2 \vec{q^{\prime}}}{2\pi} 
 e^{-B(|\vec{q^{\prime}}|^2+ \vec{q^{\prime}} \cdot \vec{q} +t/2)} F_d(4 t^{\prime})\,,
\label{eq:cs_sum2}
\end{eqnarray}
where $B \approx 6$ GeV$^{-2}$~\cite{H1:2006uea} 
is the slope of the $t$ dependence
of the diffractive $\gamma^{\ast} + N \to X +N$ differential cross section.
Note that the deuteron quadrupole form factor does not contribute to the summed cross section, see Ref.~\cite{Franco:1969qj}.

Equation~(\ref{eq:cs_sum2}) has a transparent physical interpretation. The first line is the impulse approximation (IA), where the overall factor of two reflects the contributions of two nucleons (proton and neutron) in the deuteron and the factor of $F_d(4t)$ 
originates from the interference of the proton and neutron contributions at small $t$. 
Note that the slope of the $t$ dependence of this term is twice as large as that for the coherent cross section on the deuteron in IA.
In the $t \to 0$ limit, 
$d\sigma_d^{\rm sum}(t)/dt=4(d\sigma_{\rm el}(t)/dt+d\sigma_{\rm diss}(t)/dt)$.
The second and third lines in Eq.~(\ref{eq:cs_sum2}) give the negative shadowing correction, whose strength is driven by 
the $(d\sigma_{\rm diff}/dt)/\sigma_{\rm tot}$ ratio and whose $t$ dependence is determined by the geometry of the photon-deuteron scattering 
in transverse plane, i.e.,
by the deuteron form factor and the slopes $B_{el}$ and $B$.

In LTA, the shadowing correction in $J/\psi$ photoproduction can be expressed in terms of the nuclear shadowing factor for 
the gluon distribution in the deuteron $g_d(x,Q^2)$ at small $x$. The latter is given by the following relation~\cite{Frankfurt:2011cs},
  \begin{eqnarray}
 S_g(x)&=&\frac{g_d(x,Q^2)}{2 g_N(x,Q^2)} \nonumber\\
 &=&1-\frac{d\sigma_{\rm diff}/dt(t=0)}{\sigma_{\rm tot}}
\int d t^{\prime} e^{-Bt^{\prime}} F_d(4 t^{\prime}) \,,
\label{eq:S_g}
\end{eqnarray}
where $g_N(x,Q^2)$ is the free nucleon gluon distribution. Note that all involved cross sections implicitly depend on energy (Bjorken $x$).

Equation~(\ref{eq:S_g}) allows us to rewrite the expression for the summed cross section in the following final 
compact form
\begin{eqnarray}
\frac{d\sigma_d^{\rm sum}(t)}{dt}&=&2 \left(1+F_d(4 t)\right)\left(\frac{d\sigma_{\rm el}(t)}{dt}+\frac{d\sigma_{\rm diss}(t)}{dt}\right)  \nonumber\\
&-&8\frac{d\sigma_{\rm el}(t=0)}{dt} (1-S_g(x)) e^{-(B_{\rm el}+B) t/2} \,,
\label{eq:cs_sum3}
\end{eqnarray}
where we neglected the numerically unimportant $e^{-B \vec{q^{\prime}} \cdot \vec{q}}$ term in Eq.~(\ref{eq:cs_sum2}).

Since the deuteron is the lightest nucleus and 
the STAR kinematics corresponds to $W \approx 25$ GeV~\cite{STAR:2021wwq}, 
$x=M_{J/\psi}^2/W^2 \approx 0.015$, and $Q^2 \approx M_{J/\psi}^2/4 \approx 2.4-3$ GeV$^2$ ($M_{J/\psi}$ is the mass of $J/\psi$ and $Q$ is a 
generic value of the resolution scale probed in $J/\psi$ photoproduction), the gluon shadowing correction in this case
is very small, $S_g(x)=0.99$. i.e., on the order of 1\%~\cite{Frankfurt:2011cs}.
This explicit numerical estimate supports a similar conclusion in Ref.~\cite{STAR:2021wwq}.

%
\section{LTA predictions for $t$ dependence of $d+Au$ UPC cross section and comparison to STAR data}
\label{sec:lta}

\begin{figure}[t]
\begin{center}
\epsfig{file=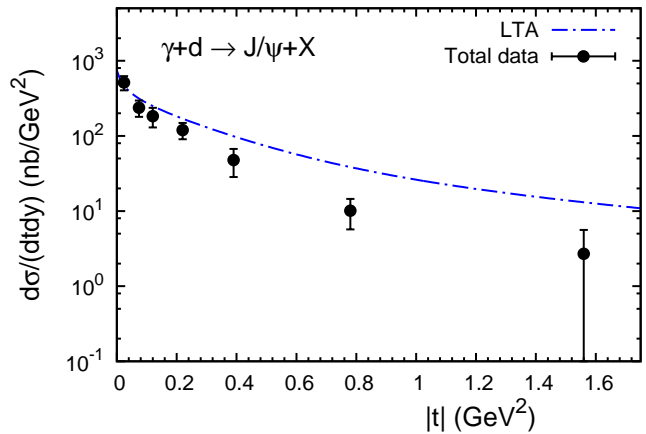,scale=1.}
\caption{The summed $\gamma +d \to J/\psi+X$ differential cross section as a function of $|t|$: the LTA predictions (blue dot-dashed curve) and the STAR data~\cite{STAR:2021wwq} for the total cross section.
The experimental statistical and systematic errors have been added in quadrature.} 
\label{fig:JPsi_RHIC}
\end{center}
\end{figure}

It is customary to present UPC cross sections as a function of the rapidity of the produced final state, in the considered case, 
the $J/\psi$ rapidity $y$. In the STAR notation, the UPC cross section and the cross section at the photon level are 
related via the photon flux $\Phi_{T, \gamma}$
\begin{equation}
\frac{d^2 \sigma^{d+Au \to J/\psi+X}}{dt dy}=\Phi_{T, \gamma} \frac{d^2 \sigma^{\gamma+d \to J/\psi+X}}{dt dy} \,,
\label{eq:upc_star}
\end{equation}
where $t =-|\vec{p}_T|^2$ and $\vec{p}_T$ is the $J/\psi$ transverse momentum.
Note that $J/\psi$ photoproduction on the gold nuclei can be safely neglected because the photon flux due to the deuteron ions is suppressed
approximately by the factor of $1/(79^2) =1/6240$ compared to that due to the gold ions.
Since the STAR results are given for $d^2 \sigma^{\gamma+d \to J/\psi+X}/(dt dy)$, our predictions can be readily
compared to the data. This is presented in Fig.~\ref{fig:JPsi_RHIC}, which shows predictions of 
Eq.~(\ref{eq:cs_sum3}) for the
summed $\gamma +d \to J/\psi+X$ cross section as a function of $|t|$ (blue dot-dashed curve) and the STAR data~\cite{STAR:2021wwq} 
for the total cross section (filled circles). The error bars represent statistical and systematic errors added in quadrature.
 A comparison of the LTA prediction to the filled circles shows that while the theory 
describes well the data within experimental errors at small $|t|$, it significantly overestimates the data for $|t| > 0.2$ GeV$^2$.

This discrepancy cannot be attributed to the gluon nuclear shadowing since it affects the summed cross section only at the 
level of a few percent. In fact, with $1-2$\% precision, Eq.~(\ref{eq:cs_sum2}) for $|t| > 0.2$ GeV$^2$ can be
written in the following approximate simple form
\begin{equation}
\frac{d\sigma_d^{\rm sum}(t)}{dt} \approx 2 \frac{d\sigma_{\rm el}(t)}{dt}+2 \frac{d\sigma_{\rm diss}(t)}{dt} \,,
\label{eq:cs_sum2_app}
\end{equation}
thus emphasizing that the gluon nuclear shadowing or details of the nuclear structure (deuteron wave function)
do not play a role here.

At this point it is important to discuss the input for the elastic and proton-dissociative cross sections
of $J/\psi$ photoproduction on the nucleon. In our analysis, we used the low-energy (LE) H1 parametrization based on the data covering the $25 \ {\rm GeV} < W < 80$ GeV interval~\cite{Alexa:2013xxa}.
In particular, we combined results of the fit for the $W$ dependence in Table 3 of Ref.~\cite{Alexa:2013xxa}
with that for the $t$ dependence in Table 2. The resulting parameterization reads
\begin{eqnarray}
\frac{d\sigma_{\rm el}(t)}{dt} &=& N_{el} \left(\frac{W}{90}\right)^{\delta_{el}}  b_{el} e^{- b_{el}|t|} \,, \nonumber\\
\frac{d\sigma_{\rm diss}(t)}{dt} &=& N_{pd} \left(\frac{W}{90}\right)^{\delta_{pd}}  \frac{f_{pd}(t)}{\int dt^{\prime}  f_{pd}(t^{\prime})} \,,\label{eq:sigma_el}
\end{eqnarray}
where $N_{el}=81 \pm 3$ nb, $\delta_{el}=0.67 \pm 0.03$, and 
$b_{el}=4.3 \pm 0.02$ GeV$^{-2}$ for the elastic case; $N_{pd}=62 \pm 12$ nb,  $\delta_{pd}=0.42 \pm 0.05$, and 
 $f_{pd}(t)=1/(1+b_{pd}|t|/n)^n$ with $n=3.58$ and $b_{pd}=1.6 \pm 0.2$ GeV$^{-2}$ for the nucleon-dissociative case.
 
Note that the H1 analysis included the proton-dissociative masses in the range $m_p < M_Y < 10$ GeV~\cite{Alexa:2013xxa}.
A simple estimate in the case of the RHIC kinematics shows that the nucleon-dissociative final state covers the rapidity interval 
$\Delta y =(1/2)\ln[(M_Y^2 +|t|)/(m_p^2+|t|)] \approx 2 $, which is adjacent to the $|y_N|=\ln[2 E_N/(m_p^2+|t|)^{1/2}] \approx 5$ 
rapidity of the final-state nucleon scattered quasi-elastically.
When comparing our calculations to the RHIC UPC data, 
one has to keep in mind that in selection of UPC events, the STAR detector covers a wide range of rapidities vetoing 
events with dissociation.  This includes $|y| <1$ covered by the Time Projection Chamber (TPC), $3.4 < |y| < 5$ covered by 
two Beam-Beam Counters (BBCs), and Zero Degree Calorimeters (ZDCs) tagging forward neutrons.
Therefore, the nucleon-dissociative state $Y$ is predominantly produced in the rapidity region covered by the STAR detector and, hence, 
can be effectively rejected.

\begin{figure}[t]
\begin{center}
\epsfig{file=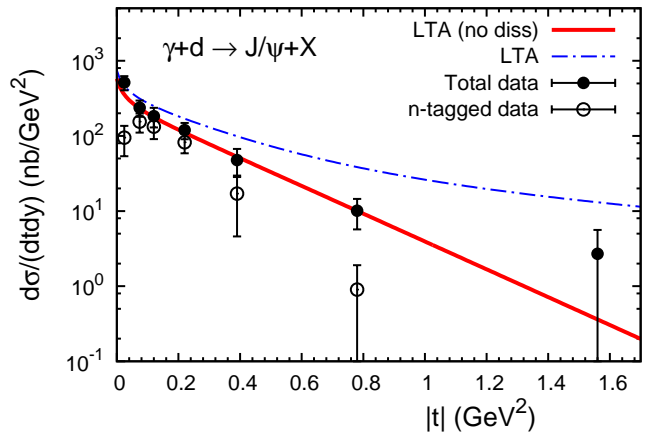,scale=1.0}
\caption{
The summed $\gamma +d \to J/\psi+X$ differential cross section as a function of $|t|$: the LTA predictions 
with the $d\sigma_{\rm diss}/dt=0$ scenario (red solid curve) vs.~the STAR data~\cite{STAR:2021wwq} for the total cross section (filled
circles). 
The data for the neutron-tagged cross section are shown by open circles.
Our result from Fig.~\ref{fig:JPsi_RHIC} is given by the blue 
dot-dashed curve.
}
\label{fig:JPsi_RHIC2}
\end{center}
\end{figure}

To explore the possibility
that the range of $M_Y$ included in the STAR analysis is significantly smaller than that in the H1 case, 
we set  $d\sigma_{\rm diss}/dt=0$ in Eq.~(\ref{eq:cs_sum3}).
Note that since $d\sigma_{\rm diss}/dt \gg d\sigma_{\rm el}/dt$ for sufficiently large $|t|$, the discrepancy mentioned 
above should originate from modeling of $d\sigma_{\rm diss}/dt$.
The resulting prediction is given by the red solid curve in Fig.~\ref{fig:JPsi_RHIC2} and leads to a very good agreement with the
STAR data over the entire range of $t$. For comparison, we also show our result from Fig.~\ref{fig:JPsi_RHIC} by the blue 
dot-dashed curve, which includes the nucleon-dissociative events.

Note that the H1 measurement of proton-dissociative $J/\psi$ photoproduction~\cite{Alexa:2013xxa} is the only collider 
data in the relevant range of $W=25$ GeV; the earlier ZEUS data~\cite{ZEUS:2002vvv} correspond to higher $W$, 
$80 \ {\rm GeV} < W < 120$ GeV.

The comparison of the LTA results to the RHIC data presented in Fig.~\ref{fig:JPsi_RHIC2} should be taken with some degree of caution.
First, the RHIC data corresponds to the rapidity range $|y| < 1$, where the UPC cross section of $J/\psi$ photoproduction 
at RHIC energies has a rapid $y$ dependence, see, e.g.~\cite{Strikman:2005ze,Frankfurt:2006tp}. 
Second, to validate our hypothesis of the absence of the nucleon-dissociative contribution, it is important 
to model the distribution over the momentum of the recoiled nucleons and to
utilize the ``n-tagged data" (open circles in Fig.~\ref{fig:JPsi_RHIC2}),
which is beyond the scope of the present article.

To explain the latter point, we notice that the incoherent cross section is given by the difference between the summed and 
coherent cross sections, which in IA reads
\begin{eqnarray}
\frac{d\sigma_d^{\rm inc}(t)}{dt} &=& \frac{d\sigma_d^{\rm sum}(t)}{dt}-\frac{d\sigma_d^{\rm coh}(t)}{dt} \nonumber\\
&=&
2 \frac{d\sigma_{\rm el}(t)}{dt}\left(1+F_d(4 t)-2|F_d(t)|^2\right) \nonumber\\
&+&2 \frac{d\sigma_{\rm diss}(t)}{dt}\left(1+F_d(4 t)\right) \,.
\label{eq:cs_incoh}
\end{eqnarray}
Thus, our hypothesis of the absence of a noticeable nucleon-dissociative contribution seems to be supported 
by ``n-tagged data" for $|t| < 0.1$ GeV$^2$ since the difference between the solid and open circles in Fig.~\ref{fig:JPsi_RHIC2}
can be explained by the contribution of coherent $J/\psi$ photoproduction off the deuteron.
In the $0.1 < |t| < 0.4$ GeV$^2$ range, the measured cross sections with and without neutron-tagging
agree within experimental errors and are also described by our calculations neglecting
the nucleon-dissociation contribution.
At the same time, the data at $|t| > 0.4$ GeV$^2$ require further analysis.

We discussed in the Introduction that the STAR data on $J/\psi$ photoproduction in $d+Au$ UPCs is described well within the CGC framework with nucleon shape fluctuations~\cite{Mantysaari:2019jhh}. 
We also explained above that the effect of leading twist gluon nuclear shadowing
as well as details of the deuteron structure are largely unimportant for $|t | \geq 0.2$ GeV and that
 the summed cross section on the deuteron should be reproduced sufficiently accurately by approximate Eq.~(\ref{eq:cs_sum2_app}).
 Therefore, the difference in predictions of the LTA and CGC approaches 
 could come from different extrapolations of the  
 $d\sigma_{\rm el}(t)/dt$ and $d\sigma_{\rm diss}(t)/dt$ cross sections on the nucleon from HERA to RHIC energies.
 Note also that the CGC approach considers cross sections integrated over $M_Y$, which makes it challenging 
 to do a comparison with the data containing various cuts on $M_Y$.

\section{Summary}
\label{sec:summary}

In this 
paper,
we argue that the STAR data on $J/\psi$ photoproduction in 
$d+Au$ UPCs 
for $t \neq 0$ primarily probes elastic and probably nucleon-dissociative $J/\psi$ photoproduction on quasi-free nucleons
and is essentially insensitive to nuclear modifications of the gluon distribution in the deuteron.  
Exploring the possibility that the nucleon-dissociative 
contribution is negligibly small, we obtain a very good description of the $t$ dependence of the STAR total cross section.
We point out that the difference between the LTA and CGC predictions 
could come from different extrapolations of the  
 elastic and nucleon-dissociative $J/\psi$ photoproduction cross sections on the nucleon from HERA to RHIC energies.

\acknowledgments
The authors would like to thank Zhoudunming Tu for stimulating our interest in this work, discussion of the STAR results, and
useful comments on an earlier draft of this letter.
The research of M.S.~was supported by the US Department of Energy Office of Science, Office of Nuclear Physics under 
Award No.~DE-FG02-93ER40771. M.S.~thanks Theory Division of CERN for
hospitality while this work was done.


\begin{thebibliography}{99}

\bibitem{Baltz:2007kq}
A.~J.~Baltz, G.~Baur, D.~d'Enterria, L.~Frankfurt, F.~Gelis, V.~Guzey, K.~Hencken, Y.~Kharlov, M.~Klasen and S.~R.~Klein, \textit{et al.}
Phys. Rept. \textbf{458}, 1-171 (2008)
[arXiv:0706.3356 [nucl-ex]].

\bibitem{Klein:2019qfb}
S.~R.~Klein and H.~M\"antysaari,
Nature Rev. Phys. \textbf{1}, no.11, 662-674 (2019)
[arXiv:1910.10858 [hep-ex]].

\bibitem{Guzey:2020ntc}
V.~Guzey, E.~Kryshen, M.~Strikman and M.~Zhalov,
Phys. Lett. B \textbf{816}, 136202 (2021)
[arXiv:2008.10891 [hep-ph]].

\bibitem{Guzey:2013xba}
V.~Guzey, E.~Kryshen, M.~Strikman and M.~Zhalov,
Phys. Lett. B \textbf{726}, 290-295 (2013)
[arXiv:1305.1724 [hep-ph]].

\bibitem{Guzey:2013qza}
V.~Guzey and M.~Zhalov,
JHEP \textbf{10}, 207 (2013)
[arXiv:1307.4526 [hep-ph]].

\bibitem{Mantysaari:2017dwh}
H.~M\"antysaari and B.~Schenke,
Phys. Lett. B \textbf{772}, 832-838 (2017)
[arXiv:1703.09256 [hep-ph]].

\bibitem{Mantysaari:2020axf}
H.~M\"antysaari,
Rept. Prog. Phys. \textbf{83}, no.8, 082201 (2020)
[arXiv:2001.10705 [hep-ph]].

\bibitem{STAR:2021wwq}
M.~Abdallah \textit{et al.} [STAR],
Phys. Rev. Lett. \textbf{128}, no.12, 122303 (2022)
[arXiv:2109.07625 [nucl-ex]].

\bibitem{Guzey:2018tlk}
V.~Guzey, M.~Strikman and M.~Zhalov,
Phys. Rev. C \textbf{99} (2019) no.1, 015201
[arXiv:1808.00740 [hep-ph]].

\bibitem{Lacombe:1981eg}
M.~Lacombe, B.~Loiseau, R.~Vinh Mau, J.~Cote, P.~Pires and R.~de Tourreil,
Phys. Lett. B \textbf{101} (1981), 139-140

\bibitem{Bauer:1977iq}
T.~H.~Bauer, R.~D.~Spital, D.~R.~Yennie and F.~M.~Pipkin,
Rev. Mod. Phys. \textbf{50}, 261 (1978)
[erratum: Rev. Mod. Phys. \textbf{51}, 407 (1979)]

\bibitem{Glauber:1955qq}
R.~J.~Glauber,
Phys. Rev. \textbf{100}, 242-248 (1955)

\bibitem{Gribov:1968jf}
V.~N.~Gribov,
Sov. Phys. JETP \textbf{29}, 483-487 (1969)

\bibitem{Frankfurt:2011cs}
L.~Frankfurt, V.~Guzey and M.~Strikman,
Phys. Rept. \textbf{512}, 255-393 (2012)
[arXiv:1106.2091 [hep-ph]].

\bibitem{Alexa:2013xxa}
C.~Alexa \textit{et al.} [H1],
Eur. Phys. J. C \textbf{73} (2013) no.6, 2466
[arXiv:1304.5162 [hep-ex]].

\bibitem{H1:2006uea}
A.~Aktas \textit{et al.} [H1],
Eur. Phys. J. C \textbf{48}, 749-766 (2006)
[arXiv:hep-ex/0606003 [hep-ex]].

\bibitem{Franco:1969qj}
V.~Franco and R.~J.~Glauber,
Phys. Rev. Lett. \textbf{22}, 370-374 (1969)


\bibitem{ZEUS:2002vvv}
S.~Chekanov \textit{et al.} [ZEUS],
Eur. Phys. J. C \textbf{26}, 389-409 (2003)
[arXiv:hep-ex/0205081 [hep-ex]].



\bibitem{Strikman:2005ze}
M.~Strikman, M.~Tverskoy and M.~Zhalov,
Phys. Lett. B \textbf{626}, 72-79 (2005)
[arXiv:hep-ph/0505023 [hep-ph]].

\bibitem{Frankfurt:2006tp}
L.~Frankfurt, M.~Strikman and M.~Zhalov,
Phys. Lett. B \textbf{640}, 162-169 (2006)
[arXiv:hep-ph/0605160 [hep-ph]].


\bibitem{Mantysaari:2019jhh}
H.~M\"antysaari and B.~Schenke,
Phys. Rev. C \textbf{101}, no.1, 015203 (2020)
[arXiv:1910.03297 [hep-ph]].





\end{thebibliography}
\end{document}